\documentclass{article}

    \PassOptionsToPackage{numbers, compress}{natbib}

\usepackage[preprint]{neurips_2023}

\usepackage[utf8]{inputenc} 
\usepackage[T1]{fontenc}    
\usepackage{hyperref}       
\usepackage{url}            
\usepackage{booktabs}       
\usepackage{amsfonts}       
\usepackage{nicefrac}       
\usepackage{microtype}      
\usepackage{xcolor}         
\usepackage{graphicx} 

\title{Improving International Climate Policy via Mutually Conditional Binding Commitments (Track 2)}

\author{
Jobst Heitzig \\
FutureLab on Game Theory and Networks of Interacting Agents\\ 
Complexity Science Department\\
Potsdam Institute for Climate Impact Research \\
\texttt{jobst.heitzig@pik-potsdam.de}
\And
Jörg Oechssler \\
Alfred Weber Institute for Economics\\
University of Heidelberg
\And
Christoph Pröschel \\
Technical University of Berlin \\
\texttt{c.proeschel@campus.tu-berlin.de}
\And
Niranjana Ragavan
\And
Yat Long Lo \\
Dyson Robot Learning Lab \\
\texttt{richie.lo@dyson.com}
}

\begin{document}

\maketitle

\begin{abstract}
  The Paris Agreement, considered a significant milestone in climate negotiations, has faced challenges in effectively addressing climate change due to the unconditional nature of most Nationally Determined Contributions (NDCs). This has resulted in a prevalence of free-riding behavior among major polluters and a lack of concrete conditionality in NDCs. To address this issue, we propose the implementation of a decentralized, bottom-up approach called the Conditional Commitment Mechanism. This mechanism, inspired by the National Popular Vote Interstate Compact, offers flexibility and incentives for early adopters, aiming to formalize conditional cooperation in international climate policy. In this paper, we provide an overview of the mechanism, its performance in the AI4ClimateCooperation challenge, and discuss potential real-world implementation aspects. Prior knowledge of the climate mitigation collective action problem, basic economic principles, and game theory concepts are assumed.
\end{abstract}

\section{Introduction}

Even though the Paris Agreement is often framed as a milestone in climate negotiations, it still proved ineffective \cite{paris}.
This is in part due to the unconditionality of most Nationally Determined Contributions (NDCs), leading to persistent free-riding by large polluters.
Only very few NDCs make use of the conditionality enabled by the Paris Agreement, and mostly only in an inconcrete or incommensurable and thus largely inconsequential way, even though the literature has demanded more conditionality \cite{victor}.

We propose to formalize this type of conditional cooperation using a decentralized, bottom-up \textit{Conditional Commitment Mechanism} that was developed earlier by one of us \cite{heitzigSSRN}, inspired by the National Popular Vote Interstate Compact \cite{NPVIC}, and was shown to have appealing theoretical properties such as flexibility and incentives for first movers.

Here we summarize the mechanism, its performance in the AI4ClimateCooperation challenge, and discuss some real-world implementation aspects. 
We assume familiarity with the real-world collective action problem of climate mitigation and some basic knowledge of economic and game-theoretic concepts as described in the challenge's white-paper \cite{zhang2022ai}.

\section{Proposed solution: The Conditional Commitment Mechanism}



The \textbf{theoretical basis} of our contribution is the {\em Conditional Commitment Mechanism} introduced in \cite{heitzigSSRN}: When players in a one-shot ``base'' game with (completely or partially) ordered action spaces and preferences that fulfill certain monotonicity conditions bind themselves via so-called {\em conditional commitment functions (CCFs)} rather than choosing actions directly, then the outcomes of the strong Nash equilibria \cite{strongequilibrium} of the resulting ``commitment'' game are provably exactly the core outcomes of the underlying base game. 
In particular, these strategic equilibria are individually rational and Pareto-optimal. 
Players who can communicate upfront should thus be able to coordinate right away on a strong equilibrium that realizes a Pareto-optimal outcome.
It was also shown in \cite{heitzigSSRN} that if the base game is repeated, also players who can {\em not} communicate upfront will still converge to Pareto-optimal outcomes when using a very simple form of non-exploitable better-response learning dynamics.
This suggests that AI agents using more sophisticated reinforcement learning algorithms should converge to such good equilibria even more reliably.

Mathematically, the CCF $c_i$ that player $i$ chooses is a function that maps each action profile $a_{-i}$ of the other players to a minimal action $a_i=c_i(a_{-i})$ that player $i$ commits to realize if the other players (denoted $-i$) realize at least the actions $a_{-i}$.\footnote{%
    Even though it is formally similar to a best-response function, a CCF does not encode best responses in the underlying non-cooperative game but can rather be interpreted as offers for cooperation. It is also somewhat similar to supply function equilibria in Cournot competition \cite{supplyfunction}.} 
The mechanism then computes the (provably existing and unique) {\em largest feasible action profile} $a^\ast$ so that $a^\ast_i\le c_i(a^\ast_{-i})$ for all $i$, and all $i$ are then bound to realize actions that are at least as large as $a^\ast_i$.
\footnote{%
    The resulting equilibria can be interpreted as a program equilibrium \cite{programequilibrium} between a specific class of programs that represent CCFs.}

\begin{figure}
    \centering
    \includegraphics[width=1.0\columnwidth,trim=10 10 300 10,clip]{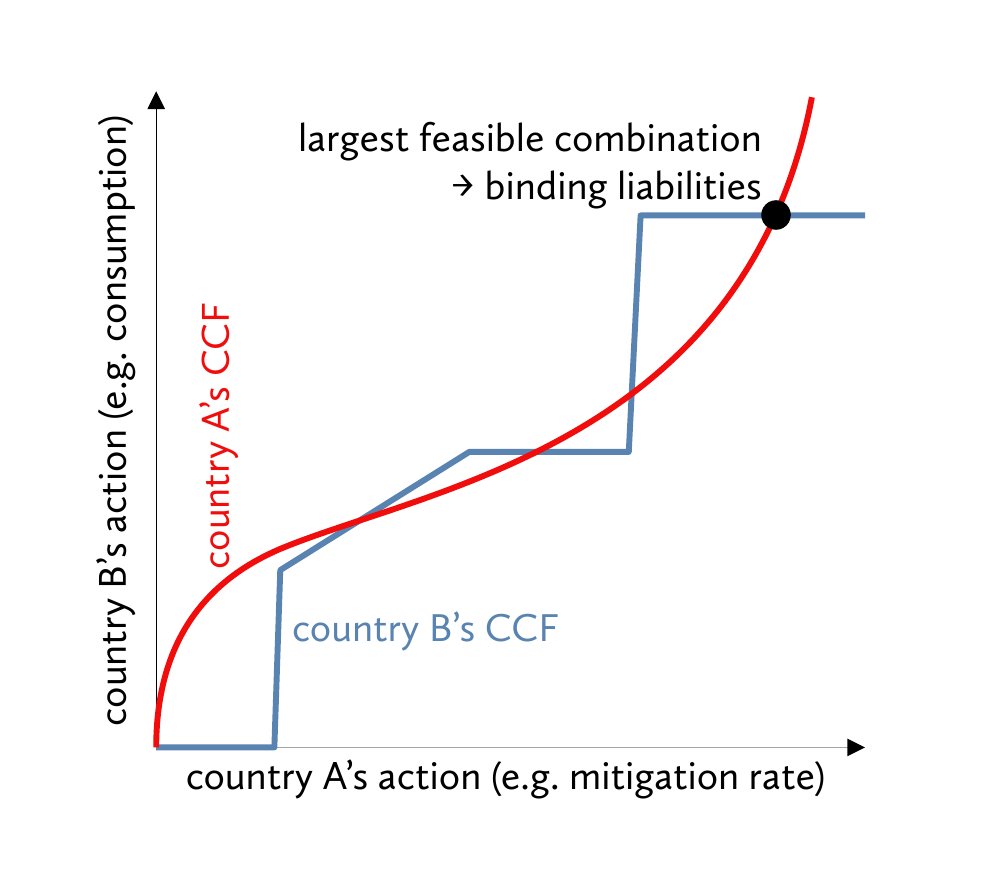}
    \caption{Two-country example of the \textbf{Conditional Commitment Mechanism}. After each country has submitted a conditional commitment function (CCF), their liabilities are computed as the the largest feasible action combination.}
    \label{fig:ccf}
\end{figure}

Since CCFs are ``action-type-agnostic'', they are a very flexible tool for making conditional offers regarding many different forms of actions (mitigation, consumption, tariffs, subsidies, contributions to funds, technology transfer, vaccine donations, military aid, whatever) and they can have any form as long as they are weakly increasing functions. 
E.g., they could mix one or more discrete steps with some linearly or nonlinearly increasing parts, etc., as illustrated in Fig.\,\ref{fig:ccf} for a case with just two countries.
For the theoretical results in \cite{heitzigSSRN}, which generalize earlier results in \cite{binary,oechssler}, already CCFs of a very simple form suffice, namely  ``two-step'' functions where $c_i(a_{-i})=a^2_i$ iff $f(a_{-i})\ge\theta^2_i$, otherwise $c_i(a_{-i})=a^1_i$ iff $f(a_{-i})\ge\theta^1_i$, and otherwise $c_i(a_{-i})=0$. 
In this, $a^2_i\ge a^1_i\ge 0$ are two offered action levels for player $i$, $f$ is some monotonic aggregation function such as the minimum or average of others' actions, and $\theta^2_i\ge\theta^1_i$ are the ``thresholds'' for this aggregate that define the condition for player $i$ to be willing to realize at least action $a^2_i$ or $a^1_i$, respectively.

Motivated by these theoretical results, we designed our \textbf{conditional commitment negotiation protocol for the RICE model} as follows:
In each of RICE's 5-year periods: (i) each country (or ``region'') $i$ simultaneously submits $k\ge 1$ 
many pairs $(a^j_i,\theta^j_i)$, $j=1\dots k$. 
(Typically, $k\le 4$ suffices for good performance.)
These encode a $k$-step CCF $c_i$ 
for this period and player.
Then (ii) the largest feasible profile $a^\ast$ is calculated and is turned into all players' ``liabilities''%
, and finally (iii) players choose economic control variables (mitigation, savings, etc.) that are constrained by these liabilities $a^\ast_i$%
.

The components of the ``action'' $a^j_i$ are from a subset 
of the following economic control variables that are defined or can be derived in the RICE model: 
{\em mitigation rate} 
(0\dots 1), 
{\em consumption rate} 
($= 1 -{}$ {\em savings rate}
), 
and {\em ``non-tariff on clean imports''}. 
The latter variable is a way to set import tariffs depending on suppliers' mitigation rates, representing a form of carbon border adjustment mechanism (CBAM). 
More specifically, the ``non-tariff on clean imports'' set by country $i$ results in the following constraint on $i$'s tariff on imports from country $j$: {\em tariff} 
$\le 1 -{}$ {\em ``non-tariff on clean imports''}${}_i$ 
$\times$ {\em mitigation rate}${}_j$. 

The rationale for this choice of variables in the RICE model is that mitigation and consumption tend to be costly for $i$ and beneficial for $-i$, and the opposite for tariffs so that players' preferences regarding these action components approximately fulfill the monotonicity requirements of the CCF theory and thus the players should converge to a strong Nash equilibrium that realizes an approximately Pareto-optimal outcome.

The thresholds $\theta^j_i$ are w.r.t.~a corresponding subset 
of the aggregate variables {\em global mitigation rate, global consumption rate, global ``non-tariff on clean imports'',} 
which are suitably weighted global averages. 
The rationale is that conditioning on global aggregates is less complex (which should accelerate learning), more flexible to fulfill (which increases the potential for cooperation), and more realistic than conditioning on individual players' actions. 

For \textbf{example}, if $k=2$, a player might offer (i) to mitigate $\ge 20\%$, consume $\ge 80\%$, and tax clean imports at rate $\le 30\%$ in the coming period if global mitigation is $\ge 30\%$, global consumption is $\ge 70\%$, and globally avg.~tariffs on clean imports are $\le 40\%$, and (ii) to even mitigate $\ge 40\%$, consume $\ge 90\%$, and tax clean imports at rate $\le 20\%$ if global mitigation is $\ge 40\%$, global consumption is $\ge 80\%$, and globally avg.~tariffs on clean imports are $\le 30\%$. Depending on whether and how this matches with other cpuntries CCFs, the player's actual economic control variables are then constrained by either (i) or (ii) or remain unconstrained for this period.

A possible \textbf{real-world implementation} of the CCF mechanism would be via unilateral legislation whereby an individual country passes a bill that (i) specifies a CCF encoding offers and conditions, and (ii) bindingly commits that country to comply with the largest action profile that is feasible (in the above sense) given all CCFs that are specified in similar bills that are put into force by other countries.

For example, country A might start by passing a bill that (i) offers to match every 2 GtC of global mitigation by $\ge 1$ GtC of domestic mitigation, up to at most 10 GtC/year, and (ii) commits country A to this action as soon as other countries have passed similar bills matching the conditions set in (i).
If then, say, countries B, C and D pass similar bills offering 8 GtC each if global mitigation is $\ge 30$ GtC, then the action profile $(10,8,8,8)$ GtC for (A,B,C,D) matches all four countries' conditions and is thus immediately becoming a binding liability without the countries having to sign any treaty or pass further legislation. 
B could later amend their bill to include a further offer to, say, contribute at least USD 10 bln to a global compensation fund if country E reduces their tariffs on certain products to $\le 10\%$. 
As soon as E passes a matching bill, both actions become additional liabilities.
This way, countries can step-wise increase their ambition regarding various interrelated variables with positive externalities for other countries in a bottom-up emerging process not requiring multilateral treaties or upfront negotiations between many governments in the usual sense.

\section{Effectiveness}

In theory, rational countries should easily realize that our mechanism allows them to cooperate effectively from the very beginning to achieve Pareto-optimal, individually and collectively rational outcomes.
They can get to this insight either by classical game-theoretical inspection of the mechanism or by reinforcement learning in a simulated environment.
In particular, our mechanism does not require a slow ramping-up of ambitions or a complex intertemporal policy involving reciprocity.
To verify this, we performed several forms of simulation experiments. 
Already in \cite{heitzigSSRN}, we verified the convergence to the core in a simple best-response dynamics in a three-player public goods game with one control variable (mitigation), linear benefits, and quadratic costs.

For this challenge, we repeated that exercise in the same simple environment with the neural-network-based reinforcement learning algorithm supplied by the competition organizers, restricted to simple ``$k$-step-function'' CCFs (see above). 
This showed convergence to fairly good outcomes significantly better than in the case without negotiations, but they did not constitute strong Nash equilibria (which can be derived easily analytically in this simple game, see Appendix). 
We suspect this is due to the insufficient capabilities of the prescribed learning algorithms (Appendix).

We noted that due to exploration, agents sometimes converge to action combinations {\em above} the Pareto-frontier, i.e., mitigating {\em more} than optimal.
To avoid this, we implemented an additional variant of the mechanism in which the ordering of the $k$ offer-condition-pairs $(a^j_i,\theta^j_i)$ play a prioritizing role, so that additional offers will not always increase ambition but might also reduce ambition to optimal levels if prioritized correctly.
Two further variations of the mechanism, as well as a temporary reward shaping trick during learning were introduced to help coordination towards a beneficial equilibrium, which in our simulations indeed tended to raise the resulting level of ambition (see Appendix for details).

Finally, we used the same shipped RL algorithm to assess performance in the RICE-N framework using several control variables.
As of the date of submission, our results in this environment are promising but mixed and somewhat inconclusive, but suggest similar issues with the RL algorithm as already identified in the simpler environment, which is why we choose not to use them as evidence in this track 2 submission even though we have submitted them to track 1. 


\section{Feasibility}

Real-world feasibility will mainly depend on the degree to which countries can enforce the necessary actions within their borders, the legal feasibility of passing the respective legislation that implements the mechanism in individual countries, which depends on the political systems of the respective countries, and on the level of bindingness of the resulting bills. 
If the latter is low, the resulting compliance problem might have to be addressed using suitable monitoring systems and additional sanctioning mechanisms, as in any form of treaty.
In many countries, citizens or NGOs could also enforce compliance with passed bills by appealing to supreme courts.
If the used CCFs are smooth enough (rather than step-functions as in the RL simulations), noncompliance could be deterred in a similar way as in ref.~\cite{LINC}.
In our simulations, we assumed full compliance.

Since in contrast to other mechanisms, ours does not require upfront multilateral negotiations between governments (even though it might profit from them) but relies purely on unilateral legislation that can be passed by parliaments, it may result in an empowerment of parliaments relative to governments and in a much larger transparency than negotiations behind closed doors, thereby also increasing democratic legitimacy. 
It might also reduce the frictions and delays of the typical two-level game of negotiation and ratification occurring in traditional treaties \cite{two-level}.

Given the salience of these benefits, this should increase incentives and possibilities for non-governmental actors (such as members of parliament, NGOs, citizens) to push for CCF legislation.
They can push for ambitious climate policy more easily when proposing conditional rather than unconditional action. 
Early moving domestic actors can argue that their country can provide an example and set the agenda and a norm by encoding certain concrete actions and conditions into its CCF. 
If early movers use linear CCFs, thereby offering to match others' ambitions with own contributions, then domestic actors in other countries can point out that now those countries' actions would have much larger leverage if they also pass a corresponding CCF.
This is similar to how charitable fundraisers sometimes incentivize donations by matching them with additional funds.
These later movers also have incentives to use metrics in their CCFs commensurable to early movers' metric in order to make their bill consequential. 

\section{Robustness}

Besides the above-mentioned possible issues with non-compliance if countries cannot bindingly commit, a general failure mode that we have not addressed is irrational behavior such as not accepting individually rational offers.
It is also unclear how bills passed under the CCF mechanism will be constrained by or influence other types of treaties that exist in parallel, such as bottom-up carbon market linkages (see, e.g., \cite{carbonmarkets}).
A coalition of countries that have established liabilities via our mechanism can however easily turn them into a formal multilateral treaty later to increase stability further, albeit losing some flexibility. 

Due to its flexibility, our mechanism can be used in a way that is robust against fluctuations in mitigation costs, economic performance etc.: If countries use smooth CCFs rather than step functions, then a country that, say, experiences a recession or unexpectedly large mitigation costs can unilaterally change their CCF and all other countries' liabilities would automatically be adjusted accordingly, without them having to amend their laws. 


\section{Ethics and climate justice}

The suggested mechanism is procedurally fair in a formal sense since it treats all countries the same. 
There is no special role for large actors. 
However, countries with a faster legislative process might gain an advantage by passing their bills earlier than others, thereby setting the agenda. 
Our theoretical results suggest that countries might converge to any outcome in the core of the base game, which might be more or less equitable. 
The fact that they also have to be individually rational for all countries can be seen as a pragmatic constraint on the level of justice that can be achieved, which is however hard to avoid in any form of free negotiations.

To push in the direction of outcomes associated with calls for Climate Justice, early movers should condition their CCFs on aggregate variables that take into account historical responsibility.
Non-country actors such as the UN, the IPCC, or NGOs, may influence this towards equity by proposing suitable aggregate metrics to condition on.

Finally, countries that already act unconditionally may be disadvantaged because they have less to offer further, which is however also a common feature of negotiations.


\section{Novelty}
The novelty of the CCF mechanism in the climate context is the switch from multilateral coordinated negotiations to unilateral binding but mutually conditional commitments, adding flexibility and speed. 

\section{Conclusion}

The proposed mechanism combines the flexibility of the Paris agreement with the bindingness and concreteness of traditional treaties. 
Between rational countries, it should converge fast to a Pareto-optimal, individually and collectively rational outcome.
Next steps are the design of robust legal language for domestic bills implementing the mechanism by key players within varying legal frameworks, the identification of specific unilateral offers and conditions to be included in a first wave of corresponding unilateral bills, and talks with relevant policy-makers that could advance these bills.

~

\subsection*{Author contributions}
J.H. designed the mechanism and performed the game-theoretical analysis. 
J.H. and C.P. implemented the mechanism in the challenge framework.
C.P. performed Monte-Carlo simulations of the learning process.
N.R. visualized the results and did research on continuous action spaces.
Y.L.L. consulted about reinforcement learning methodology.
J.O. consulted about economic theory.
J.H., C.P. and N.R. wrote and edited the paper.

\subsection*{Carbon accounting}

We roughly estimated our computations to have caused carbon emissions on the order of 1.5 metric tons of CO$_2$ equivalents.

{\small\sffamily
\bibliographystyle{plain}
\bibliography{refs} 
}

\section*{Appendix}

\subsection*{Simple additional public goods game}

In addition to RICE-N, we also performed simulations with a very simple public goods game to assess the RL algorithm's performance.
Each of the $N$ regions $i$ has constant capital and production, both normalized to 1.
There is no growth or capital accumulation or depreciation, and there is no trade (savings rates and trade-related control variables are thus ignored).
The only meaningful control variable is thus the regional {\em mitigation rate} $\mu_i\in[0,1]$, and regional emissions are then $1 - \mu_i$ per period.
We assume quadratic mitigation costs of $\gamma\mu_i^2$ which are subtracted from production, where $\gamma>0$ is a cost parameter.
The global temperature anomaly $T$ simply equals global emissions here, and there is no independent temperature dynamics.
Regional climate damages are linear, $2\beta T = 2\beta(N - \sum_j \mu_j)$, so that regional period payoffs are $\Pi_i = 1 - \gamma\mu_i^2 - 2\beta N + 2\beta \sum_j \mu_j$.
In the base game's Nash equilibrium (i.e., without negotiations), all players choose $\mu_i \equiv \mu^{NE} = \beta/\gamma$, giving regional period payoffs of $\Pi^{NE} = 1 - 2\beta N + (2 N - 1)\beta^2/\gamma$.
In contrast, the socially optimal choice is $\mu_i \equiv \mu^\ast = N\beta/\gamma$, giving higher regional period payoffs of $\Pi^\ast = 1 - 2\beta N + N^2\beta^2/\gamma$.

When putting $N=2$, $\beta=0.09$ and $\gamma=0.3$, 
we get $\mu^{NE}=0.3<\mu^\ast=0.6$ and $N\Pi^{NE}=1.442<N\Pi^\ast=1.496$. 
With the conditional commitment mechanism, both players can easily achieve the social optimum by putting $(a_i,\theta_i)\equiv(0.6,0.6)$, where $a_i=\mu_i$ and $\theta_i=\sum_j\mu_j/N$, which is a Pareto-optimal and individually and collectively rational Nash equilibrium of the mechanism.
Indeed, with the simple learning dynamics we had earlier performed in \cite{heitzigSSRN}, players converge fast to this outcome.

Still, at least in our simulations, the shipped RL algorithms from this competition were unfortunately unable to learn the optimal equilibrium even in this almost trivial two-player setup and rather ended up with global period payoffs around 1.47, i.e., about halfway between the disagreement and socially optimal outcomes.

\subsection*{Variants of the mechanism to improve learnability by RL agents}

\subsubsection*{Prioritized offers}

In the basic version of the mechanism, the pairs $(a^j_i,\theta^j_i)$ are just used to define a $k$-step CCF and the mechanism selects the largest action profile feasible in all CCFs. 
Once that is above the Pareto-frontier, players can only reduce it again to optimal levels by removing offers and replacing them by lower ones. In order to avoid falling back to zero liabilities when removing too large offers, they would first need to add a lower offer before removing the larger one. 
But since adding a lower offer makes no immediate difference but only has a delayed effect as soon as the larger offer is removed, this is hard to learn for RL agents.

We thus studied a variant of the mechanism in which the ordering of the submitted points $(a^j_i,\theta^j_i)$ plays a prioritizing role.
More precisely, the mechanism would first construct all CCFs as before and find those combinations of submitted points $(a^j_i,\theta^j_i)$ that are feasible in all CCFs. 
Rather than selecting the largest such, it would then for each player $i$ find that feasible profile which is the most prioritized by that player.
It would then select the supremum of all those ``favourite feasible'' profiles.
So if all players prioritize a point on the Pareto-frontier over the other possible agreements above the Pareto-frontier, the former would be selected, allowing players to get down to the Pareto-frontier more easily when inadvertently being above it. 
For rational (rather than learning) agents, this makes no difference, the strong equilibria are the same as before, they can only be learned more easily now.

\subsubsection*{Borda-scoring}
While the prioritizing approach improved performance somewhat in the simple public good game with few players, it did not do so when the number of players is too large, because even a single player's wrong prioritization can drive the supremum of the favourite feasible profiles above the Pareto-frontier. 
We therefore added a further modification in which taking the supremum as the last step is replaced by performing a simple form of automatic voting mechanism using the submitted prioritization. 
More precisely, we used the Borda scoring method for this. 

\subsubsection*{Option to adjust towards the mean}

Another issue for convergence is the non-uniqueness of Pareto-efficient, individually and collectively rational possible agreement points in action space and the resulting equilibrium selection and coordination problem.
To address this, we finally studied a variant of the mechanism that has an additional step in each period: 
After all countries have submitted their prioritized list of points $(a^j_i,\theta^j_i)$ for this period, each country $i$ can choose to replace their highest-priority point $(a^1_i,\theta^1_i)$ by a certain form of average of all other players' highest-priority points $(a^1_j,\theta^1_j)$.
If all countries choose to do so, they would thereby agree on this average as the outcome.

\subsection*{Suggested improvements in RL algorithms}

It was somewhat unclear to us whether the RL algorithms supplied to participants (based on the A2C algorithm) can be expected to converge to a Nash equilibrium (NE) or some refinement of NE (such as subgame-perfect or strong equilibrium), or at least to a low-regret state of the negotiation game in the first place, and if so to which particular equilibrium if there are many (as is usually the case in complex games like this).

From comparing the used RL approach with established alternative approaches to find equilibria used in environmental economics (e.g., optimization with welfare weights using gradient descent in action space), we suspect that the used RL agents might not be reliably able to find Pareto-efficient and collectively rational equilibria for at least two independent reasons.

\subsubsection*{No discretization of continuous action space}

For one, the RL agents use a randomized policy over a discretized action space rather than a deterministic policy over the actual continuous action space. 
Learning via gradient descent thus cannot exploit the topological structure of the action space. 
Rather than shifting actions smoothly around in action space, they shift probability mass from points in action space to possibly quite far away regions in action space. 
But given the largely approximately convex public-good structure of the relevant utility functions, it is not to be expected that Pareto-efficient NEs are mixed rather than pure strategies, and that a classical gradient descent in action space (rather than in probability space) should converge well. 
we hence suggest dropping the action discretization and use algorithms such as the continuous-action version of A2C \cite{continuous}.

\subsubsection*{Meta-learning MARL and anticipation}

Independently of this, the fact that the learning algorithm for the RL agent representing country $i$ is treating the RL agents representing the other countries as part of the environment makes the environment non-stationary since all RL agents are trained simultaneously and thus have changing policies.
Since convergence theorems usually have to assume stationarity, one can expect weak convergence because of this non-stationarity. 
This is not surprising given the longstanding insights from the early theory of learning in games that showed that already simple best-response dynamics can lead to oscillations rather than convergence since each player is ``chasing a moving target''. 
To overcome this, we suggest using state-of-the-art Multi-Agent Reinforcement Learning (MARL) algorithms that treat the environment and other agents differently, such as the ``meta-learning'' approach in \cite{metalearning}.

We believe that particularly designs in which an RL agent anticipates the other RL agents' next learning steps would be very conducive to convergence in a negotiation context. 
This is because if $i$ switches in a learning step from making a bad offer to making a good offer, this will only pay off after the other players have switched in their next learning step from not accepting the offer to accepting the offer.

\subsubsection*{Reward shaping via temporary altruism}

As we were not supposed to change the RL algorithms used in the competition, the only modification of the RL algorithm that we actually tested in our simulations was a certain temporary modification of the reward function in the hope of helping the simultaneously learning agents converge to good rather than bad Nash equilibria.
The idea is to replace the selfish reward function $r_i = \Pi_i$ (regional period payoff) by a partially altruistic reward function $r_i' = (1-\alpha)r_i + \alpha\sum_j r_j/N$ with an altruism parameter $\alpha\in[0,1]$.
If we set $\alpha\approx 1$ in the early phase of learning, all agents try to maximize approximately the same objective function, so there is no conflict between the learners and they should consequently converge to this maximum.
As this maximum is on the Pareto-frontier of the actual payoff space, it is either within or close to that part of the Pareto-frontier that is individually and collectively rational, i.e. in the core. 
If in the later phase of learning $\alpha$ is then slowly reduced to zero, our expectation was that the learners would then move from the found point on the Pareto frontier to a Nash equilibrium of the mechanism that is not too far away from it and still on the Pareto frontier. 
Since all points in the core are indeed Nash equilibria, the hope was that thus the learners would converge to the core rather than to worse equilibria.

Despite this reasoning, this modification did not show a significant improvement of final reward in our experiments in which we tested several different ``schedules'' for $\alpha$, all having $\alpha\equiv 0$ during the last third of the learning process. 
This however can also be due to the possible general issues with convergence to Nash equilibrium that we suspected (see above).
We, therefore, believe that this reward shaping by temporary altruism during learning should be revisited with more suitable RL algorithms.


\end{document}